\documentclass[twocolumn,english,aps,prl,superscriptaddress]{revtex4-1}
\usepackage[latin9]{inputenc}
\pdfoutput=1
\usepackage{amsmath}
\usepackage{amssymb}
\usepackage{graphicx}
\usepackage{esint}

\makeatletter
 
 \@ifundefined{textcolor}{}
 {%
   \definecolor{BLACK}{gray}{0}
   \definecolor{WHITE}{gray}{1}
   \definecolor{RED}{rgb}{1,0,0}
   \definecolor{GREEN}{rgb}{0,1,0}
   \definecolor{BLUE}{rgb}{0,0,1}
   \definecolor{CYAN}{cmyk}{1,0,0,0}
   \definecolor{MAGENTA}{cmyk}{0,1,0,0}
   \definecolor{YELLOW}{cmyk}{0,0,1,0}
 }

\makeatother

\usepackage{babel}
\begin{document}

\title{Enhanced quantum nonlinearities in a two mode optomechanical system}

\author{Max Ludwig}

\email{max.ludwig@physik.uni-erlangen.de}

\affiliation{Institute for Theoretical Physics, Universität Erlangen-Nürnberg,
Staudtstr. 7, 91058 Erlangen, Germany}

\author{Amir H. Safavi-Naeini}

\affiliation{Thomas J. Watson, Sr, Laboratory of Applied Physics, California Institute
of Technology, Pasadena, California 91125, USA.}

\author{Oskar Painter}

\affiliation{Thomas J. Watson, Sr, Laboratory of Applied Physics, California Institute
of Technology, Pasadena, California 91125, USA.}

\author{Florian Marquardt}

\affiliation{Institute for Theoretical Physics, Universität Erlangen-Nürnberg,
Staudtstr. 7, 91058 Erlangen, Germany}

\affiliation{Max Planck Institute for the Science of Light, Günther-Scharowsky-Straße
1/Bau 24, 91058 Erlangen, Germany}
\begin{abstract}
In cavity optomechanics, nanomechanical motion couples to a localized
optical mode. The regime of single-photon strong coupling is reached
when the optical shift induced by a single phonon becomes comparable
to the cavity linewidth. We consider a setup in this regime comprising
two optical modes and one mechanical mode. For mechanical frequencies
nearly resonant to the optical level splitting, we find the photon-phonon
and the photon-photon interactions to be significantly enhanced. In
addition to dispersive phonon detection in a novel regime, this offers
the prospect of optomechanical photon measurement. We study these
QND detection processes using both analytical and numerical approaches. 
\end{abstract}
\maketitle
\textit{Introduction. -} By coupling mechanical resonators to the
light of optical cavities the emerging field of optomechanics \cite{2009_FM_ReviewOptomechanics}
aims at observing quantum mechanical behavior of macroscopic systems.
The ultimate goal is the regime where single phonons and photons interact
strongly. New architectures and progress in design and fabrication
pave the way towards realizing strong coupling even at the single-photon
level in optomechanical systems \cite{2008_Murch_Observation_nature,2008_Brennecke_CavityOptomechanics,2010_Purdy_TunableCavityOptomechanics,2011_Teufel_SidebandCooling_Nature,2011_Chan_LaserCoolingNanomechOscillator,2012_Verhagen_QuantumCoherentCoupling}.
This development has stimulated several theoretical works that analyze
the generic optomechanical system, i.e. a single optical mode coupled
to a single mechanical mode, in the regime of strong coupling. Non-classical
effects are found in the dynamics of the mechanical resonator \cite{2008_ML_OptomechInstab,2011_Nunnenkamp_SinglePhotonOptomechanics_PRL,2011_Qian_QuantumSignatures}
and the statistics of the light field \cite{2011_Rabl_PhotonBlockade_PRL,2011_Nunnenkamp_SinglePhotonOptomechanics_PRL,2012_Kronwald_FCS}
if the photon-phonon coupling rate $g_{0}$ becomes comparable to
both the decay rate of the cavity $\kappa$ and the mechanical oscillation
frequency $\Omega$.

In this paper, we show how an optomechanical setup consisting of two
optical modes coupled to a mechanical resonator \cite{Thompson_2008,2010_Grudinin_PhononLaserAction,2011_Safavi-Naeini_PPT}
can be brought into a novel regime that significantly enhances the
size of the quantum nonlinearity. We derive an effective Hamiltonian
of the system that captures the regime of strong single-photon optomechanical
coupling $g_{0}/\kappa\gtrsim1$ and large mechanical frequencies.
In our analysis the difference between optical level splitting and
mechanical frequency, $\delta\Omega=2J-\Omega$, appears as a crucial
parameter. It enters the coupling rate $g_{0}^{2}/\delta\Omega$ that
characterizes the coherent interaction among photons and between photons
and phonons. If this dispersive optical frequency shift exceeds the
cavity decay rate, one enters what we will call the strong dispersive
coupling regime: $g_{0}^{2}\gtrsim\kappa\delta\Omega$. Since $\delta\Omega$
can be made much smaller than $\Omega$, this condition is easier
to achieve than the corresponding one for the generic optomechanical
system, $g_{0}^{2}\gtrsim\kappa\Omega$. This is relevant in particular
because optomechanical systems have by now reached the regime of large
mechanical frequencies, see for example \cite{2011_Chan_LaserCoolingNanomechOscillator,2011_Teufel_SidebandCooling_Nature,2012_Verhagen_QuantumCoherentCoupling},
where they are less susceptible to thermal fluctuations and optomechanical
cooling is more efficient.

\begin{figure}[ht]
 \includegraphics[width=1\columnwidth]{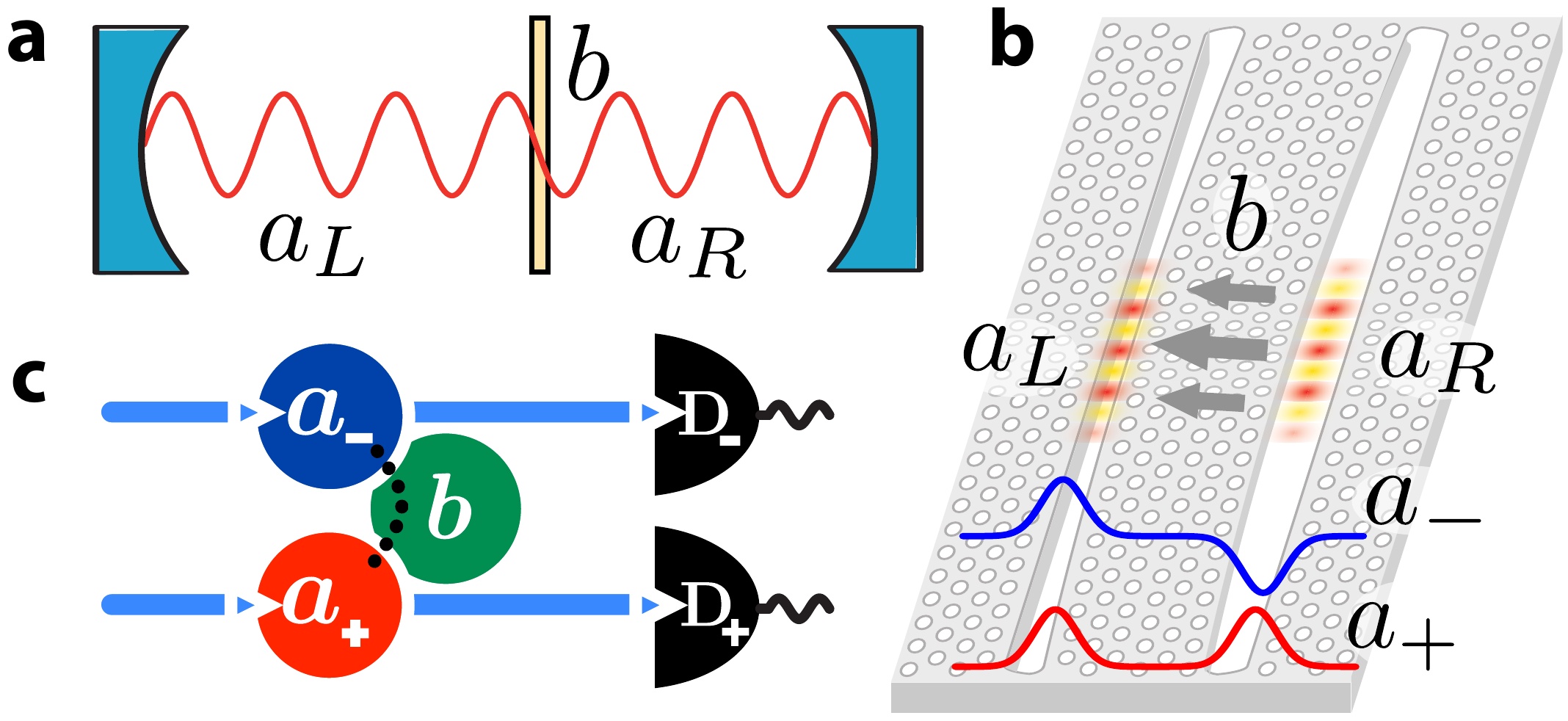}

\caption{(a),(b) Example implementations of the double cavity setup for enhanced
quantum nonlinearities: membrane in the middle (a) and optomechanical
crystal setup (b). (c) Scheme depicting the mechanical mode ($b$)
and the optical modes $a_{\pm}$. For the photon and phonon detection
applications discussed in this paper, the cavities are assumed to
be driven by independent laser sources and the transmitted signal
is measured by photodetectors $D_{\pm}$.}

\label{Fig1} 
\end{figure}

As a first application of the enhanced phonon-photon interaction we
investigate the possibility of a quantum non-demolition (QND) detection
of the phonon number. A measurement of this kind has been proposed
in a pioneering work by Thompson \textit{et al.} \cite{Thompson_2008}
for a setup where a dielectric membrane is placed inside an optical
cavity. Subsequently, this QND scheme \cite{2008_Jayich_DispersiveOptomechanics_NJP,2009_Miao_SQLForProbingMechEnergyQuant,2011_Gangat_PhononNumberQuJumps,2011_Lambert_MacrorealismInequality}
and other features of such a two mode system \cite{2008_Bhattacharya_PartiallyReflectiveMirrors,2009_Zhao_ThreeModeOptoacousticParamp,2010_Nunnenkamp_CoolingAndSqueezingQuadraticCoupling,2011_Cheung_NonadiabaticOptomechanicalHamiltonian,2011_Biancofiore_MembraneAbsorption,2011_Wu_LandauZenerPhononLasing}
have been studied in detail. An increase of the nonlinear coupling
by making use of the full spectrum of cavity modes has been demonstrated
in \cite{2010_Sankey_StrongAndTunable,2011_Hill_MechanicalTrapping,2011_Karuza_TunableLinearAndQuadraticCoupling}.
However, the analysis has so far been restricted to cases, where the
influence of individual photons is weak. Furthermore, it was assumed
that the mechanical and optical timescales separate. Hence the previous
analysis did not capture the enhancement of the optomechanical nonlinearity,
which, as we show below, results in an increased read-out rate.

As a completely new feature of optomechanical systems, our effective
description reveals strong photon-photon interaction for mechanical
frequencies comparable to the optical mode splitting. As we show below,
this interaction opens up the possibility of a QND measurement of
the photon number. The two mode optomechanical system can therefore
be assigned to a larger class of optical systems whose ultimate goal
is the realization of QND photon detection on the level of single
quanta \cite{1998_Grangier_QNDMeasurementsInOptics}.

In our analysis of the phonon and photon Fock state measurements we
discuss the limitations due to quantum noise and confirm our predictions
by numerical simulations of the dissipative quantum dynamics.

\textit{Model. -} We consider an optomechanical setup consisting of
two optical modes ($a_{\pm}$, frequencies $\omega_{\pm}$) and one
mechanical mode ($b$, frequency $\Omega$) that is described by a
Hamiltonian 
\begin{eqnarray}
H & = & H_{0}+H_{{\rm int}}+H_{{\rm drive}}+H_{{\rm diss}},\label{eq:Hamiltonian-basic}\\
H_{0} & = & \hbar\omega_{-}a_{-}^{\dagger}a_{-}+\hbar\omega_{+}a_{+}^{\dagger}a_{+}+\hbar\Omega b^{\dagger}b\label{eq:Hamiltonian-basic2}\\
H_{{\rm int}} & = & -\hbar g_{0}(b^{\dagger}+b)(a_{+}^{\dagger}a_{-}+a_{-}^{\dagger}a_{+}\big)\label{eq:Hamiltonian-basic3}\\
H_{{\rm drive}} & = & \hbar\alpha_{\pm}(e^{i\omega_{L\pm}t}a_{\pm}+H.c.)\label{eq:Hamiltonian-basic4}
\end{eqnarray}
The optomechanical coupling rate is denoted by $g_{0}$, and both
optical modes are pumped by laser sources at rates $\alpha_{\pm}$.
The optical cavities are characterized by the photon decay rates into
the reflection channel ($\kappa_{\pm,r}$) and into the transmission
channel ($\kappa_{\pm,t}$) with $\kappa_{\pm}=\kappa_{\pm,r}+\kappa_{\pm,t}$.
We assume that the transmitted signal from each of the modes can be
filtered and measured independently using a photodetector ($D_{\pm}$),
see Fig. \ref{Fig1}(c). The mechanical resonator couples to a thermal
bath at a rate $\Gamma$ with a bath occupation given by $n_{{\rm th}}$.
In the following, we assume the mechanical frequency to be high enough
and the bath temperature to be low enough such that the oscillator
is sufficiently close to the ground state.

A Hamiltonian of the form of Eq. (\ref{eq:Hamiltonian-basic}) is
found both in the ``membrane in the middle''-setup \cite{Thompson_2008},
in coupled microtoroid resonators \cite{2010_Grudinin_PhononLaserAction}
and in optomechanical crystals \cite{2011_Safavi-Naeini_PPT}. The
optical modes $a_{\pm}$ constitute normal modes $a_{\pm}=(a_{L}\pm a_{R})/\sqrt{2}$,
where $a_{L,R}$ denotes geometrically distinct modes with an original
Hamiltonian $\tilde{H}=\tilde{H}_{0}+\tilde{H}_{{\rm int}}$, where

\begin{equation}
\tilde{H}_{{\rm int}}=-\hbar J(a_{L}^{\dagger}a_{R}+H.c.)-\hbar g_{0}(b^{\dagger}+b)(a_{L}^{\dagger}a_{L}-a_{R}^{\dagger}a_{R})\label{eq:H_mim}
\end{equation}
and $\tilde{H}_{0}=\hbar\omega(a_{L}^{\dagger}a_{L}+a_{R}^{\dagger}a_{R})+\hbar\Omega b^{\dagger}b$.
The frequency splitting of the normal modes is thus given by the photon
tunnel coupling rate $J$, $\omega_{-}-\omega_{+}=2J$.

\begin{figure}[t]
\includegraphics[width=1\columnwidth]{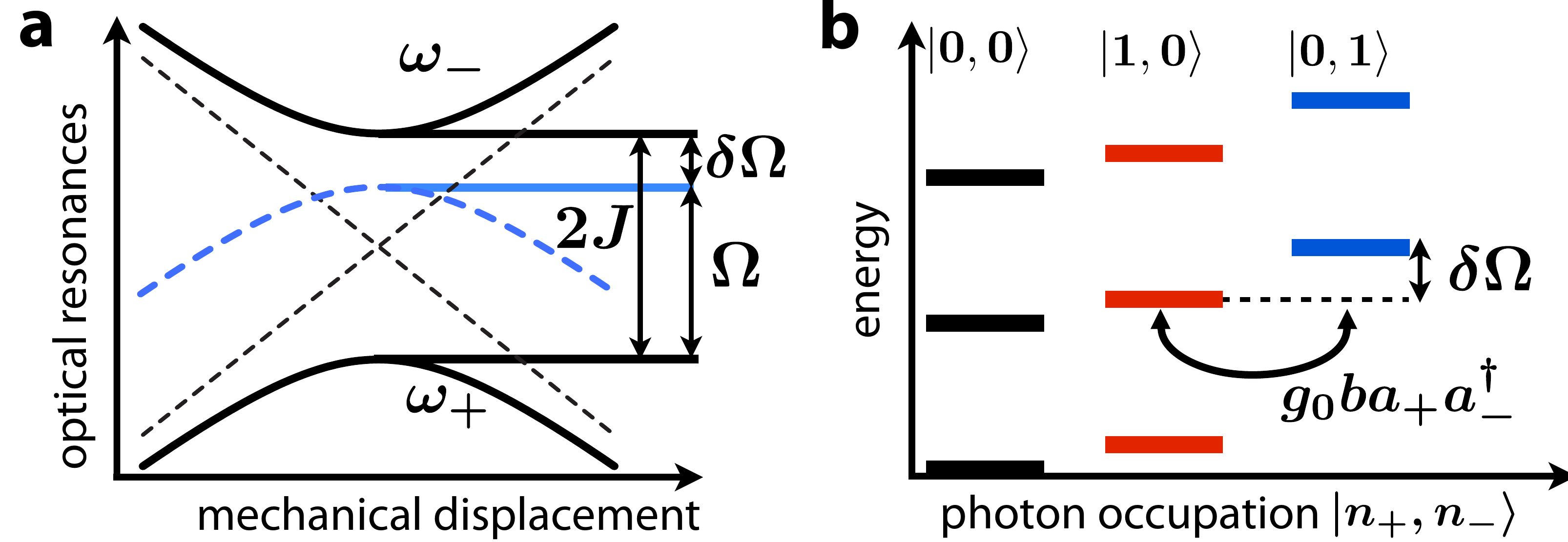}

\caption{(a) Optical resonances as a function of mechanical displacement. For
$\delta\Omega=2J-\Omega\ll\Omega,J$, the regime of enhanced effective
quantum nonlinearity is reached. (b) Energy level scheme of the double
cavity optomechanical system. The most relevant second-order transition
process is indicated.}

\label{Fig1B} 
\end{figure}

In the approach of \cite{Thompson_2008,2008_Jayich_DispersiveOptomechanics_NJP,2009_Miao_SQLForProbingMechEnergyQuant,2011_Gangat_PhononNumberQuJumps}
the optical resonances are calculated as $\omega\pm\sqrt{J^{2}+(g_{0}\tilde{x})^{2}}\approx\omega_{\pm}\pm\frac{g_{0}^{2}}{2J}\tilde{x}^{2}$
(see Fig. \ref{Fig1B}(a)), where $\tilde{x}=b^{\dagger}+b$ is the
mechanical displacement in units of the mechanical ground state width
and where it is assumed that $J\gg g_{0}\tilde{x}$. Note that $\tilde{x}$
is treated as a quasi-static variable (in the sense of the Born-Oppenheimer
approximation, with photons playing the role of electrons). This approach
therefore has to fail if the optical frequency splitting and the mechanical
excitation energy become comparable.

\textit{Effective Description.$ $ -} The effect of the optomechanical
interaction to first order in $g_{0}$ can be readily described in
the following picture. A photon initially placed in the left (or right)
cavity mode starts oscillating between the left and right part of
the cavity at a frequency $2J$: $\big(a_{L}^{\dagger}a_{L}-a_{R}^{\dagger}a_{R}\big)(t)\approx a_{+}^{\dagger}(0)a_{-}(0)e^{-2iJt}+H.c.+\mathcal{O}(g_{0}^{2})$.
Accordingly, the radiation pressure force $F=g_{0}\sqrt{2\hbar m\Omega}\big(a_{+}^{\dagger}a_{-}+a_{-}^{\dagger}a_{+})$
varies sinusoidally in time. This force drives mechanical oscillations
$x_{{\rm osc}}=F/[m(\Omega^{2}-4J^{2})]$ and $p_{{\rm osc}}=\frac{1}{\Omega^{2}-4J^{2}}F'(t)$,
where $F'(t)=-2iJ\big(a_{+}^{\dagger}a_{-}-a_{-}^{\dagger}a_{+}\big)(t)$. 

To take these elementary dynamics into account, we shift the oscillator
by $x_{osc}$ and $p_{osc}$ via a unitary transformation $H_{{\rm eff}}=e^{iS}(H_{0}+H_{{\rm int}})e^{-iS}$,
with $S=x_{{\rm osc}}p/\hbar+p_{{\rm osc}}x/\hbar$. This procedure
exactly eliminates the interaction to first order in $g_{0}$ and
results in an effective Hamiltonian

\begin{eqnarray}
H_{{\rm eff}} & = & H_{0}+\hbar\frac{g_{0}^{2}}{2}\Big(\frac{1}{2J-\Omega}+\frac{1}{2J+\Omega}\Big)\Big(n_{-}-n_{+}\Big)\Big(b^{\dagger}+b\Big)^{2}\nonumber \\
 &  & +\hbar\frac{g_{0}^{2}}{2}\Big(\frac{1}{2J-\Omega}-\frac{1}{2J+\Omega}\Big)\big(a_{+}^{\dagger}a_{-}+a_{+}a_{-}^{\dagger}\Big)^{2},\label{eq:Heff_0}
\end{eqnarray}
where $n_{\pm}=a_{\pm}^{\dagger}a_{\pm}$ and where we disregard terms
of order $g_{0}^{3}/\delta\Omega^{2}$. In the limit of vanishing
tunnel coupling, $J\to0$, the unitary transformation reduces to a
shift of the mechanical position due to a static radiation pressure
force. In this case the effective Hamiltonian is given by $H_{0}-\hbar g_{0}^{2}/\Omega(a_{L}^{\dagger}a_{L}-a_{R}^{\dagger}a_{R})^{2}$
in correspondence to the ``polaron transformation'' for the generic
single-mode setup \cite{1997_Mancini_PonderomotiveControl,1997_Bose_PreparationOfNonclassicalStates,2011_Rabl_PhotonBlockade_PRL,2011_Nunnenkamp_SinglePhotonOptomechanics_PRL}.
The most interesting regime is entered if the mechanical frequency
becomes comparable to the optical splitting, i.e. $\delta\Omega=2J-\Omega\ll J,\Omega$:

\begin{eqnarray}
H_{{\rm eff}} & = & H_{0}+\hbar\frac{g_{0}^{2}}{\delta\Omega}\big(n_{+}n_{-}+n_{-}+n_{-}n_{b}-n_{+}n_{b}\big)\label{eq:Heff}
\end{eqnarray}
where $n_{b}=b^{\dagger}b$ and where we neglect terms of the order
$g_{0}^{2}/(2J+\Omega)$ and rapidly rotating terms like $b^{\dagger2}$,
$(a_{+}^{\dagger}a_{-})^{2}$.

\textit{Phonon detection. - }The effective Hamiltonian of Eq. (\ref{eq:Heff_0})
enables us to discuss optomechanical QND phonon detection in its most
general form, going beyond previous discussions \cite{Thompson_2008,2008_Jayich_DispersiveOptomechanics_NJP,2009_Miao_SQLForProbingMechEnergyQuant,2011_Gangat_PhononNumberQuJumps}.
The optical frequencies are shifted by $\mp g_{0}^{2}(\frac{1}{2J-\Omega}+\frac{1}{2J+\Omega})n_{b}$.
We note that in the limit $\Omega\ll J$ the result of \cite{2008_Jayich_DispersiveOptomechanics_NJP}
is recovered. However, for mechanical frequencies comparable to the
optical splitting, i.e. $\delta\Omega=2J-\Omega\ll2J$, the frequency
shift per phonon $\delta\omega=g_{0}^{2}/\delta\Omega$ is greatly
enhanced. We stress that the enhancement of the frequency shift is
observable even in the weak coupling regime $g_{0}\ll\kappa_{\pm}$,
where the cavity modes have to be strongly driven in order to detect
the transmission phase shift in a homodyne measurement \cite{Thompson_2008,2011_Gangat_PhononNumberQuJumps}.
In the following, however, we focus on the regime where both $\Omega\approx2J$
and $g_{0}\gtrsim\kappa_{\pm}$ and where single quanta affect the
optical and mechanical modes strongly.

\begin{figure}[ht]
\includegraphics[width=1\columnwidth]{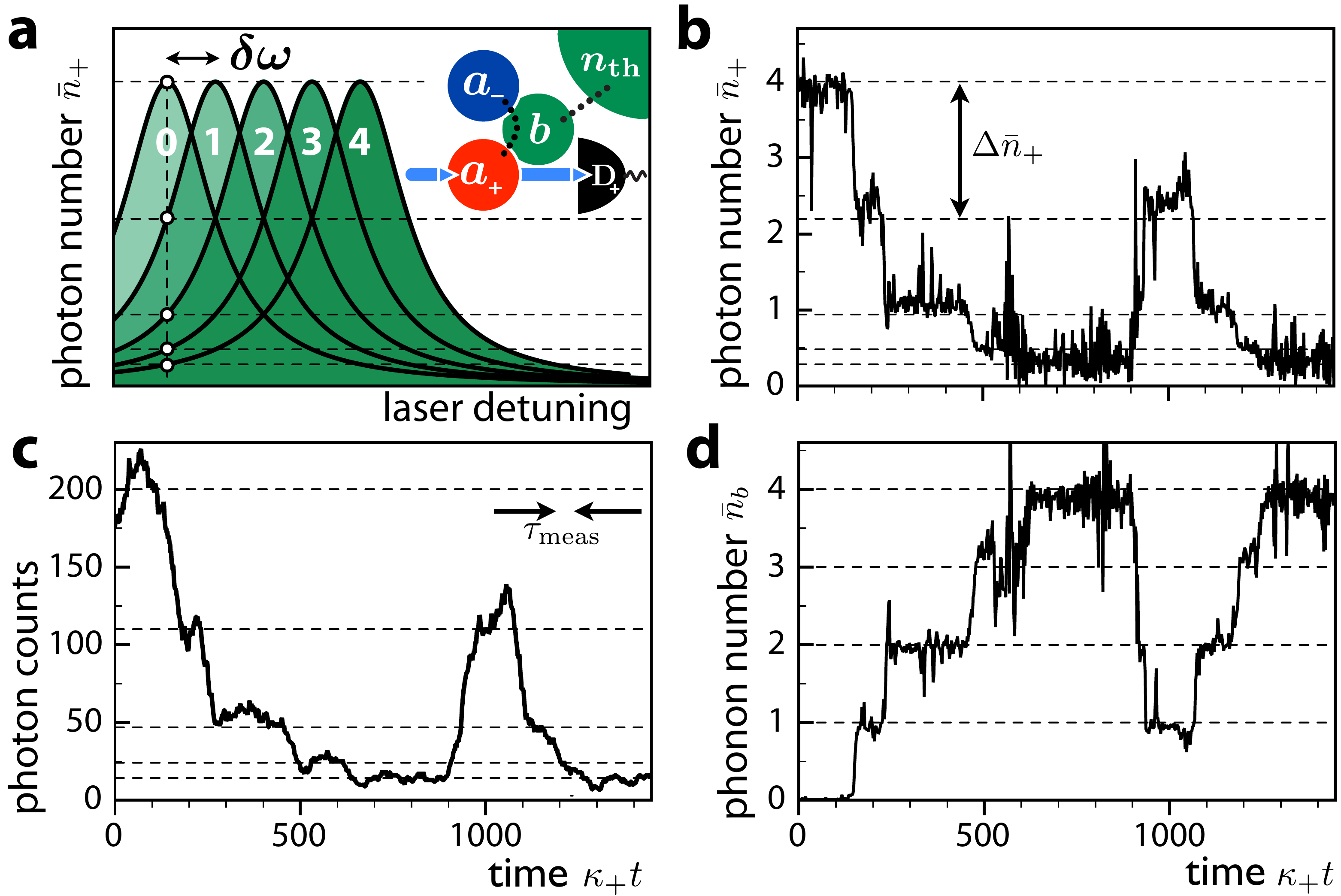}

\caption{Phonon detection in the weak dispersive coupling regime $\delta\omega=g_{0}^{2}/\delta\Omega<\kappa_{+}$,
for single-photon strong coupling $g_{0}/\kappa_{+}=3$: (a) Schematic
illustration of the resonances of the detection mode corresponding
to phonon number states 0,1,2,3,4. A jump between phonon Fock states
can be detected if a difference $\Delta n_{+}$ in the intracavity
photon number is resolved. (b),(d) Quantum trajectories of the photon
number in the detection mode, $\bar{n}_{+}=\langle a_{+}^{\dagger}a_{+}\rangle$
(b), and the phonon number, $\bar{n}_{b}=\langle b^{\dagger}b\rangle$
(d) from a numerical simulation of the stochastic master equation.
$\delta\Omega=20\kappa_{+}$, $n_{{\rm th}}=2$, $\Gamma=10^{-3}\kappa_{+}$,
$\alpha_{+}=\kappa_{+}$, $\omega_{L+}=\omega_{+}$, $\kappa_{-}=10^{-2}\kappa_{+}$,
$\kappa_{\pm,t}=0.9\kappa_{\pm}$. (c) Photon counts recorded at the
photodetector $D_{+}$ within an interval $[t-\tau_{{\rm meas}},t]$,
where a measurement time of $\tau_{{\rm meas}}=50\kappa_{+}^{-1}$
considerably smaller than the lifetime of a phonon state was chosen
($\Gamma_{{\rm th}}^{-1}\approx140\kappa_{+}^{-1},\,\Gamma_{{\rm ind}}^{-1}\approx1100\kappa_{+}^{-1}$).}

\label{Fig2a} 
\end{figure}

The experimental protocol for detecting the phonon number is to pump
one of the optical modes (here $a_{+}$) with a laser at frequency
$\omega_{L+}$ and measure the transmitted signal using a photodetector
($D_{+}$). The second mode ($a_{-}$) is undriven, playing the role
of an idle spectator (though it will become important for dissipative
processes, see below). We first study the spectrum of the detection
mode $a_{+}$, i.e. the photon number $\bar{n}_{+}$ as a function
of detuning $\omega_{+}-\omega_{L+}$. In steady state, the spectrum
consists of several resonances with spacing $\delta\omega$ corresponding
to different phonon number states. In a situation where the optical
frequency shift per phonon $\delta\omega$ is smaller than the cavity
linewidth $\kappa_{+}$, the resonances overlap, see Fig. \ref{Fig2a}(a).
In the following section, we will discuss this weak
dispersive coupling regime (even though $g_{0}/\kappa$
will still be taken on the order of one). Note that the strong dispersive
regime is also relevant, both for phonon and photon detection, and
we will come back to it when discussing photon measurements. The time
evolution of the mechanical state can be monitored by pumping the
detection mode at fixed detuning and recording the photon counts at
the detector during an interval $\tau_{{\rm meas}}$. A quantum jump
in the phonon number changes the number of intracavity photons by
$\Delta\bar{n}_{+}$ and, accordingly, the number of detected photons
by $\kappa_{+,t}\Delta\bar{n}_{+}\tau_{{\rm meas}}$. The shift in
photon number can be estimated as $\Delta\bar{n}_{+}\approx\bar{n}_{+}\delta\omega/\kappa_{+}$,
where we disregard a prefactor that depends on the detuning. The measurement
time $\tau_{{\rm meas}}$ has to be chosen large enough, such that
the measured signal exceeds the photon number uncertainty, i.e. $\Delta\bar{n}_{+}\kappa_{+,t}\tau_{{\rm meas}}>\sqrt{\bar{n}_{+}\kappa_{+,t}\tau_{{\rm meas}}}$
\cite{2008_ClerkDevoretGirvinFMSchoelkopf_RMP} or equivalently: 
\begin{equation}
\tau_{{\rm meas}}>\frac{\kappa_{+}^{2}/\kappa_{+,t}}{\delta\omega^{2}\bar{n}_{+}}.\label{eq:meas_time}
\end{equation}
On the other hand, the measurement time has to be smaller than the
lifetime of a phonon Fock state which is governed by thermal fluctuations
at rate $\Gamma_{{\rm th}}$ and by decoherence induced via the optical
modes at rate $\Gamma_{{\rm ind}}$: 
\begin{equation}
{\rm max}\big(\Gamma_{{\rm th}},\Gamma_{{\rm ind}}\big)\:\tau_{{\rm meas}}<1.\label{eq:meas_cond}
\end{equation}
The thermalization rate of the phonon state $\bar{n}_{b}$ is given
by $\Gamma_{{\rm th}}=\Gamma\big((n_{{\rm th}}+1)\bar{n}_{b}+n_{{\rm th}}(\bar{n}_{b}+1)\big)$
in the uncoupled system. The major contribution to $\Gamma_{{\rm ind}}$
stems from the process where a phonon is annihilated while a photon
tunnels from the $a_{+}$ to the $a_{-}$ mode and decays. A calculation
according to Fermi's golden rule yields $\Gamma_{{\rm ind}}\approx g_{0}^{2}\bar{n}_{+}\bar{n}_{b}\kappa_{-}/\delta\Omega^{2}$.
It follows that single-photon strong coupling, i.e. $g_{0}^{2}>\kappa_{+}\kappa_{-}$,
is required to obtain a signal to noise ratio bigger than one, as
has already been shown by \cite{2009_Miao_SQLForProbingMechEnergyQuant}
for the limiting case of small mechanical frequencies $\Omega\ll J$.
We note that a phonon measurement using the $a_{-}$ mode for detection
can be described analogously, the main qualitative difference being
that the cavity-induced decoherence processes excite phonons and potentially
cause an instability.

To simulate the envisaged QND phonon measurement, we employ the Lindblad
master equation for the system's density matrix $\rho$, 
\begin{equation}
\frac{d}{dt}\rho=-i[H,\rho]/\hbar+\sum_{{\rm unobserved}}\mathcal{D}[c_{i}]\rho+\sum_{{\rm observed}}\mathcal{D}[d_{i}]\rho\label{eq:SME}
\end{equation}
 where $\mathcal{D}[A]\rho=A\rho A^{\dagger}-\frac{1}{2}A^{\dagger}A\rho-\frac{1}{2}\rho A^{\dagger}A$.
The unobserved channels are the photon decay into the reflection channels
$c_{1,2}=\sqrt{\kappa_{\pm,r}}a_{\pm}$ and the coupling between the
mechanical resonator and the thermal environment with $c_{3}=\sqrt{\Gamma(n_{{\rm th}}+1)}b$
and $c_{4}=\sqrt{\Gamma n_{{\rm th}}}b^{\dagger}$, while the transmission
channels $d_{1,2}=\sqrt{\kappa_{\pm,t}}a_{\pm}$ are under observation.
We unravel the time evolution into quantum jumps \cite{1998_Plenio_QuJumpRMP}
$\rho(t+dt)=d_{i}\rho(t)d_{i}^{\dagger}/\langle d_{i}^{\dagger}d_{i}\rangle(t)$
that occur with probability $p_{i}(t)=\gamma_{i}\langle d_{i}^{\dagger}d_{i}\rangle(t)dt$,
and into the deterministic part $\rho(t+dt)=\rho(t)-(i[H,\rho(t)]/\hbar-\sum_{i}\mathcal{D}[c_{i}]\rho(t))dt+\sum_{i}\{\gamma d_{i}^{\dagger}d_{i}/2,\rho(t)\})dt$
plus subsequent normalization. A quantum jump with $d_{1,2}=\sqrt{\kappa_{\pm,t}}a_{\pm}$
is interpreted as a detection event at the photodetector $D_{+}$
or $D_{-}$, respectively. Figure \ref{Fig2a} (b)-(d) shows trajectories
from such a simulation. The phonon number jumps between the Fock states
$0$ and $4$, driven by thermal fluctuations (Fig. \ref{Fig2a}d).
The photon number in the detection mode follows the time evolution
of the mechanical mode (Fig. \ref{Fig2a}b). Thus, by monitoring the
photon counts at the photodetector (Fig. \ref{Fig2a}c) a QND measurement
of the phonon number is achieved. In contrast to earlier numerical
analysis \cite{2011_Gangat_PhononNumberQuJumps}, our results apply
to the general case of a two-sided cavity and thereby confirm the
limits imposed by quantum noise \cite{2009_Miao_SQLForProbingMechEnergyQuant}.
Moreover, they show the strong enhancement of the coupling in the
design considered here.

\textit{Photon detection. - }As a novel feature of the system, we
identify the dispersive photon-photon interaction in the effective
Hamiltonian (\ref{eq:Heff}). We note that the interaction term vanishes
in the limit of small mechanical frequencies $\Omega\ll J$ and therefore
did not appear in previous works. Here we demonstrate the prospects
of a QND measurement of the photon number $n_{+}$ using the $a_{-}$
mode for detection. The roles of the two optical modes are chosen
as to suppress the influence of unwanted transitions from the $a_{-}$
mode to the energetically lower-lying $a_{+}$ mode. Both modes are
driven independently by a laser and the data from the photodetector
$D_{-}$ is used to extract the information about the photon number
$n_{+}$. We assume that the detection mode has a lower finesse than
the signal mode, i.e. $\kappa_{-}\gg\kappa_{+}$, such that a sufficiently
large number of photons arrives at the detector $D_{-}$ while the
state of $a_{+}$ is only weakly perturbed by the photons in $a_{-}$.

In the weak dispersive coupling regime, $g_{0}^{2}<\kappa_{-}\delta\Omega$,
we find a required measurement time of 
\begin{equation}
\tau_{{\rm meas}}>\frac{\kappa_{-}^{2}/\kappa_{-,t}}{\delta\omega^{2}\bar{n}_{-}}\label{eq:tau_meas_2}
\end{equation}
with a frequency shift per photon of $\delta\omega=g_{0}^{2}/\delta\Omega$,
in analogy to the case of phonon detection discussed above (see also
Fig. \ref{Fig2a}). In order to detect the photon state $\bar{n}_{+}$
within its lifetime, it is also required that $\tau_{{\rm meas}}<1/\bar{n}_{+}\kappa_{+}$.
Moreover, the measurement would be spoiled if a phonon were to be
excited during the measurement time, since $a_{-}$ actually measures
$n_{+}+n_{b}$. We therefore demand that both the thermalization rate
$\Gamma_{{\rm th}}$ and the rate for the optically induced heating
process, given by $g_{0}^{2}\bar{n}_{-}\kappa_{+}/\delta\Omega^{2}$,
are smaller than the measurement rate $\tau_{{\rm meas}}^{-1}$. From
the latter condition it follows that single-photon strong coupling,
i.e. $g_{0}^{2}/\kappa_{+}\kappa_{-}>1$, is also required for an
undisturbed photon detection.

\begin{figure}[t]
\includegraphics[width=1\columnwidth]{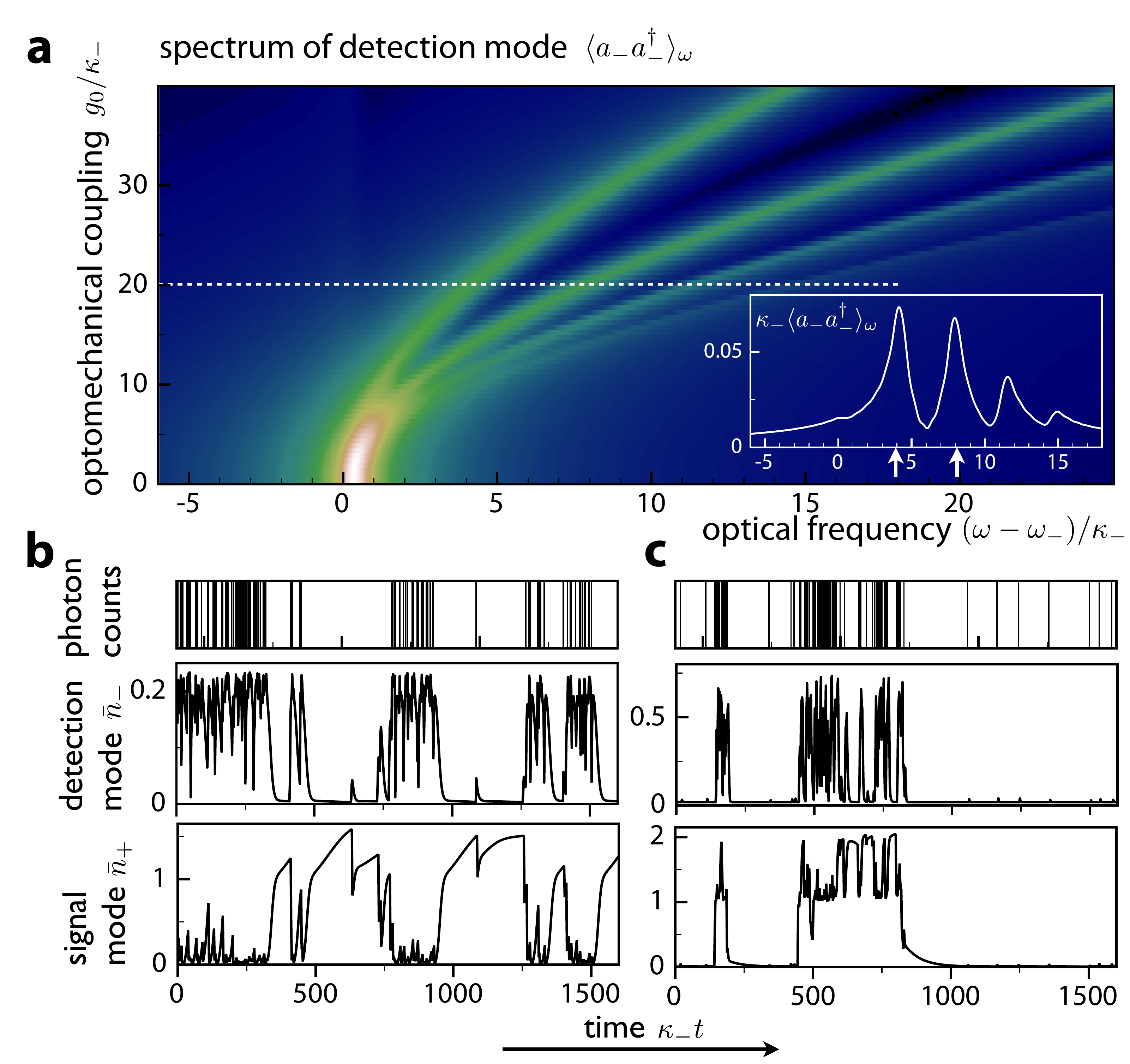}

\caption{(a) Spectrum of the detection mode, $\langle a_{-}a_{-}^{\dagger}\rangle_{\omega}=\int e^{i\omega\tau}\langle a_{-}(t+\tau)a_{-}(t)^{\dagger}\rangle d\tau$,
in the presence of a strongly driven signal mode, $\bar{n}_{+}=1$.
With increasing optomechanical coupling rate $g_{0}$, the splitting
between the resonance peaks grows like $\delta\omega=g_{0}^{2}/\delta\Omega$.
The inset shows the spectrum for $g_{0}=20\kappa_{-}$ (cut indicated
in main figure). (b),(c) Quantum trajectories for the detection mode
driven at the (b) zero-photon resonance, $\omega_{L-}-\omega_{-}=\delta\omega$,
and (c) at the one-photon resonance, $\omega_{L-}-\omega_{-}=2\delta\omega$.
This clearly shows the anti-correlation or correlation, respectively,
between signal and detection modes induced by the photon interaction.
$\delta\Omega=100\kappa_{-}$, $\kappa_{+}=10^{-2}\kappa_{-}$, $\kappa_{\pm,t}=0.9\kappa_{\pm}$,
$\omega_{L+}=\omega_{+}$, $\alpha_{L-}=\kappa_{-}/4$, $\alpha_{L+}=\kappa_{+}/2$$ $,
$n_{{\rm th}}=0$, $\Gamma=\kappa_{-}$. }

\label{Photon_detection} 
\end{figure}

In the strong dispersive regime, $g_{0}^{2}>\kappa_{-}\delta\Omega$,
a strong projective measurement of the photon number (or analogously
the phonon number) can be performed as illustrated in Fig. \ref{Photon_detection}.
The spectrum of the detection mode $a_{-}$, i.e. the intensity as
a function of laser detuning, shows well-resolved resonances with
spacing $\delta\omega$, see Fig.\ref{Photon_detection} (a). The
weights of the peaks correspond to the photon number distribution
of the signal mode. This is in close analogy to the theoretical and
experimental results of \cite{2006_Gambetta_Qubit-PhotonInteractions,2007_Schuster_ResolvingPhotonNumberStates}
where a qubit coupled to a microwave cavity was used to measure the
photon distribution. The quantum trajectory simulations (Fig. \ref{Photon_detection}(b),(c))
reveal strong measurement induced back-action leading to (anti-)correlation
between signal and detection mode. Whenever the photodetector $D_{-}$
registers photons from the detection mode, the state of the signal
mode $a_{+}$ is projected into the zero- or one-photon Fock state
depending on the detuning of the detection mode. This projection leads
to a disruption of the coherent evolution of the signal mode as is
clearly visible in Figs. \ref{Photon_detection}(b),(c)). We note
that in the regime $\tau_{{\rm meas}}^{-1}>\kappa_{+}$, this kind
of measurement backaction affects the quantum evolution significantly.
Indeed, it can be shown that the photons impinging on the signal mode
$a_{-}$ from the coherent laser source tend to be prevented from
entering the cavity due to the continuous observation of the photon
number inside the cavity. This is a manifestation of the Quantum Zeno
effect, as analyzed in \cite{2009_Helmer_QNDPhotoDetection}.

\emph{Experimental prospects}. Single-photon strong
coupling, i.e. $g_{0}>\kappa$, has been demonstrated in optomechanical
systems where the mechanical element is a cloud of cold atoms \cite{2008_Murch_Observation_nature,2008_Brennecke_CavityOptomechanics,2010_Purdy_TunableCavityOptomechanics}.
In principle, currently available setups of this kind are extensible
to a two-mode design by making use of the spectrum of transverse cavity
modes \cite{2010_Sankey_StrongAndTunable}. Reaching $\Omega\approx2J$
would additionally require larger trapping frequencies, $\Omega>\kappa$.

A number of optomechanical systems exhibit large mechanical frequencies
of a few ${\rm GHz}$, and $\Omega\approx2J$ has been demonstrated
\cite{2010_Grudinin_PhononLaserAction,2011_Safavi-Naeini_PPT,2011_Hill_MechanicalTrapping}.
Single-photon strong coupling, however, is yet to be reached in solid-state
systems. The current record is achieved in optomechanical crystal setups, $g_{0} \approx 0.007\, \kappa \approx 2\pi \times 1 \rm{MHz}$ \cite{2012_Chan_OptimizedOptomechanicalCrystalCavity}. Utilizing nanoslots~\cite{2005_Robinson_UltrasmallModeVolumes} to enhance the local optical field in
such structures offers the prospect of coupling rates above $10\,\text{MHz}$. 
Advances in design, fabrication and material properties are expected to lead
to high-quality optical cavities with $\kappa/2\pi\approx10\,{\rm MHz}$
\cite{2008_Tanaka_PhotonicCrystalNanocavity,2008_Notomi_Ultrahigh-QNanocavity}.
These developments, taken together, should make $g_{0} > \kappa$ attainable.

\textit{Conclusions and Outlook. - } The results presented here demonstrate
how the design flexibility of photonic crystals and other optomechanical
systems can be exploited to significantly enhance nonlinear coupling
rates, and how to benefit therefrom in the deep quantum regime. Besides
the dispersive QND measurement schemes based on the two-mode structure
addressed here, one may think of applying the enhanced photon-photon
and photon-phonon coupling for studies of optomechanical quantum many-body
effects (e.g. in arrays), or for further applications in quantum information
processing (see also the related work by Stannigel et al. \cite{2012_Stannigel_OptomechanicalQIP}).
The coherent Kerr-type interaction introduced here can form the basis
for an all-optical switch and moreover directly permits to engineer
a quantum phase gate (based on the conditional phase shift) for photonic
or phononic qubits. In addition, the mechanical degrees of freedom
can also serve as a quantum memory \cite{2011_Chang_SlowingAndStoppingLight_NJP},
and optomechanical interactions yield a quantum interface between
solid-state, optical and atomic qubits \cite{2010_Stannigel_LongDistanceQuCommunication,2011_Safavi-Naeini_PPT}.
The combination of these ingredients will make optomechanical systems
a promising integrated platform for quantum repeaters and general
``hybrid quantum networks``.

This work was supported by the DARPA/MTO ORCHID program through a
grant from the AFOSR, the DFG Emmy-Noether and an ERC starting grant,
and the Institute for Quantum Information and Matter, an NSF Physics
Frontiers Center with support of the Gordon and Betty Moore Foundation.
ML thanks OJP for his hospitality at Caltech.

\bibliographystyle{unsrt}
\bibliography{bib}

\end{document}